\documentclass[aps,prl,showpacs,twocolumn,superscriptaddress]{revtex4}
\usepackage{graphicx}
\usepackage{amsmath}
\bibstyle{apsrev.bib}

\newcommand{\beq}{\begin{equation}}
\newcommand{\eeq}{\end{equation}}

\begin{document}

\title{Scaling of a collapsed polymer globule in 2D}
\author{Marco Baiesi} 
\affiliation{Instituut voor Theoretische Fysica, K.U.Leuven, B-3001, Belgium}  

\author{Enzo Orlandini} 
\affiliation{INFM-Dipartimento di Fisica, Universit\`a di Padova,  
I-35131 Padova, Italy.}
\affiliation{Sezione INFN, Universit\`a di Padova, I-35131 Padova, Italy.}    

\author{Attilio L. Stella}
\affiliation{INFM-Dipartimento di Fisica, Universit\`a di Padova,
I-35131 Padova, Italy.}  
\affiliation{Sezione INFN, Universit\`a di Padova, I-35131 Padova, Italy.}  

\date{\today}

\begin{abstract}
Extensive Monte Carlo data analysis gives clear evidence that
collapsed linear polymers in two dimensions fall in the universality 
class of athermal, dense self-avoiding walks, as conjectured by 
B.~Duplantier [Phys.\ Rev.\ Lett.\ {\bf 71}, 4274 (1993)].
However, the boundary of the globule has self affine roughness
and does not determine the anticipated
nonzero topological boundary contribution to entropic
exponents. Scaling corrections are due to
subleading contributions to the partition function corresponding 
to polymer configurations with one end located on the
globule-solvent interface.

\end{abstract}
\date{June 28, 2005}


\maketitle

Polymers in solution are the subject of intense studies since
several decades~\cite{desCloiseauxJannink}. In the dilute case
and in good solvent (high temperature $T$)
excluded volume effects favor swollen configurations
for a long linear chain.
On the other hand, in poor solvent (low $T$) effective
attractive interactions between monomers dominate and the
typical conformations are those of a compact globule.
The transition between swollen and collapsed polymer
regimes is marked by the theta point~\cite{desCloiseauxJannink,Vanderzande}. 
A main achievement of the Coulomb gas~\cite{Nienhuis87} and
conformal invariance approaches has been the exact characterization of 
the scaling properties of polymers in the 
swollen~\cite{Nienhuis82,Duplantier_PRL86:branched} and
theta regimes~\cite{Coniglio_JMS_PRB87,DuplantierSaleur_PRL87:_exact_theta,
Vanderzande_SS_PRL91} in two dimensions (2D).
On the other hand, the situation is still far from
settled as far as the entropic scaling of the collapsed
phase is concerned.

A crucial feature expected for a collapsed globule is the
presence of a rather sharp boundary separating it from
the surrounding solvent. This led Owczareck et 
al.~\cite{Owczarek_PB_PRL93:collapsed}
to conjecture the presence of a factor growing like the exponential
of the boundary length in the statistical partition
sum describing the globule. It was then pointed out by 
Duplantier~\cite{Duplantier_PRL93:comment}
that with free boundary conditions such type of 
factor appears naturally for models of dense polymers (DP), in which a 
self avoiding walk (SAW) covers an assigned region of the lattice 
visiting a fixed fraction of
sites~\cite{Duplantier_JPA86:dense,DuplantierSaleur_NPB87:dense}. 
He also argued that a collapsed globule should
have the exactly known scaling exponents of DP 
with smooth free boundary in 2D. 
Since DP have only excluded volume effects, this conjecture
implies that collapsed configurations should not be sensibly
influenced by the attractive interactions. A further, less
obvious assumption is
that the globule boundary acts as a simple, smooth perimeter
confining the polymer~\cite{Owczarek_PB_PRL93:reply}.
So far, the exponents conjectured in this way have never been 
confirmed by numerical
investigations~\cite{Bennet-Wood_BGOP_JPA94,Nidras_JPA96:lowT}.

In this Letter we show that the entropic scaling of
collapsed polymers in 2D is consistent with the universality class
of DP. Our analysis elucidates geometrical properties of the 
globule-solvent interface
and the role they play in determining
exponents and strong finite size corrections to the 
asymptotic behaviors.

\begin{figure}[!b]
\includegraphics[angle=0,width=6.8cm]{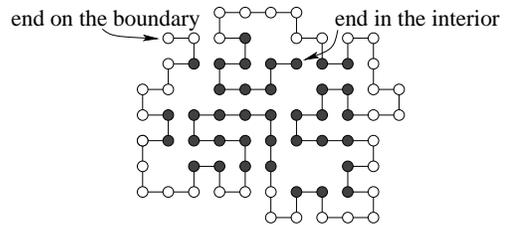}
\caption{Collapsed SAW, with sites on the boundary (empty circles)
distinguished from sites in the interior (dark circles).\label{fig:1}}
\end{figure}

Let us consider linear SAW's $w$ of $|w|=N$ steps 
on square lattice~\cite{Vanderzande} (Fig.~\ref{fig:1}). 
If to each pair of nearest neighbor
sites visited not consecutively by the SAW (contact)
is associated an attractive potential energy $-\epsilon$ ($\epsilon > 0)$,
the model undergoes the theta collapse transition upon varying
$T$. This transition can be monitored
from the asymptotic behavior of the partition function
$Z_N=\sum_{|w|=N} \exp[\frac{\epsilon}{T}  C(w)]$,
where the sum extends to all possible $w$ with an end at a
fixed origin, and $C(w)$ is
the number of nearest neighbor contacts in $w$. 
Below the theta temperature $T_{\Theta}$, 
$Z_N$ is expected \cite{Owczarek_PB_PRL93:collapsed} 
to have the following asymptotic behavior:
\beq 
Z_N (T) \simeq A\,\mu(T)^N \mu_1(T)^{\sqrt N} N^{\gamma - 1}
\label{eq:Zdense}
\eeq
where $A$ is an amplitude,
$ \mu(T) $ is a bulk free energy per step depending also on lattice structure,
and $\mu_1(T) < 1$ is a boundary term, associated with the existence
of a sharp interface separating the typically
globular region occupied by the SAW from the rest of the lattice.
The factor $\mu_1^{\sqrt N}$
implies a boundary contribution to the dimensionless total free energy,
$\ln(Z_N)$. Indeed, $\sqrt N$ is the average number of SAW steps 
on the boundary  of the globule, under the plausible 
assumption that this boundary has a fractal 
dimension equal to $1$. If one restricts the sum in Eq.~(1) 
to walks which start and end
at the origin (polygons), 
the resulting partition ($Z_N^0$) has the same asymptotics 
as in Eq.~(1), except for a different exponent $\gamma_0$ replacing $\gamma$.
Both $\gamma$ and $\gamma_0$ could take $T$-independent
universal values in the collapsed polymer regime, as
also implied by the conjecture in Ref.~\cite{Duplantier_PRL93:comment}. 
On the basis of the analogy with 
DP~\cite{Duplantier_JPA86:dense,DuplantierSaleur_NPB87:dense}, 
Duplantier conjectured $\gamma-\gamma_0 =19/16$, and $\gamma_0
=5/6$, implying $\gamma = 97/46$~\cite{Duplantier_PRL93:comment}. 
This value of $\gamma_0$
has a purely topological interpretation and was
argued by assuming that for $N \to \infty$
the boundary of the globule becomes a smooth continuous arc without 
wedges~\cite{Duplantier_PRL93:comment}.
An analysis of extensive exact enumerations
was performed in Ref.~\cite{Bennet-Wood_BGOP_JPA94}, 
in order to determine $\gamma-\gamma_0$ from extrapolations of
$Z_N/{Z^0_N}$. Considering such a ratio one expects the
exponential and stretched exponential factors in Eq.~(1)
to simplify in numerator and denominator, leaving an 
$N$-dependence $\sim N^{\gamma - \gamma_0}$.
The estimated $\gamma -\gamma_0 \approx 0.92$
was inconsistent with the conjecture of Ref.~\cite{Duplantier_PRL93:comment}.
An effort to determine $\gamma$ was subsequently made
on the basis of extensive grand canonical Monte Carlo 
sampling~\cite{Nidras_JPA96:lowT}:
the estimate $\gamma \approx 1.09$, again appears to rule out
the conjecture of Ref.~\cite{Duplantier_PRL93:comment}.

A correct interpretation of this contradictory 
scenario, and a solution of the puzzle
can be achieved once elucidated all physical consequences of the existence
of a well defined interface between collapsed globule and solvent.
This interface does not only imply the presence of a stretched 
exponential factor in the asymptotic partition function in Eq.~(1). 
Indeed, unlike in the swollen regime, the typical configurations of a 
polymer globule can be naturally partitioned into distinct groups,
depending on the location of the chain ends with respect 
to the boundary.
This circumstance provides a natural source of scaling corrections to
the asymptotic power law scaling in Eq.~(1). 
For example, imagine one 
manages to restrict the sum in $Z_N$ to compact configurations 
in which one end of the chain is located on the globule boundary, while 
the other one falls in the interior. 
We ask how the resulting ``interior-boundary'' partition function, 
$Z_N^{ib}$, should scale compared to $Z_N$. 
The fact that scaling should not be 
influenced by attractive interactions and that the interfacial 
boundary has fractal dimension equal to $1$, suggests that a simple
$N$-dependent geometrical factor should connect the two partition 
functions: $Z_N^{ib} \sim N^{-1/2} Z_N $. This factor $N^{-1/2}$ 
represents the probability that one of the ends of a compact chain of
$N$ steps (globule area $\sim N$) ends on the globule
boundary (length $\sim N^{1/2}$), if one assumes that
a chain end falls with equal probability anywhere
within the globule. 

Of course, we also take into account
that the fraction of walks with both ends on the boundary
is in turn negligible with respect to $Z_N^{ib}$.
Indeed, we can define also a restricted partition
sum $Z_N^b$ to which only chain configurations with both
ends on the boundary contribute. In this case the factor
relating the restricted partition to $Z_N$ will be $N^{-1}$,
since for large $N$ one should regard as independent
events the occurrences of boundary locations for the two ends.
Summarizing, if we further define $Z_N^i$ as the partition 
restricted to configurations with both chain ends in the 
interior of the globule, the following asymptotics should 
be expected 
\begin{subequations}
\label{eq:ZZZ}
\begin{eqnarray}
Z_N^{i}   &\simeq& A_i\, \mu^N \mu_1^{\sqrt{N}} N^{\gamma-1} 
\label{eq:Zb}\\
Z_N^{ib} &\simeq& A_{ib}\, \mu^N \mu_1^{\sqrt{N}} N^{\gamma_{ib}-1}
\label{eq:Zbs}\\ 
Z_N^{b}   &\simeq& A_b\, \mu^N \mu_1^{\sqrt{N}} N^{\gamma_b-1}
\label{eq:Zs}
\end{eqnarray}
\end{subequations}
where $A_i$, $A_{ib}$, and $A_b$ are suitable amplitudes, 
while $\gamma_{ib} = \gamma-1/2$ and $\gamma_{b} = \gamma-1$.
Clearly the sum of the three $Z$'s above must yield
$Z_N$. So, the behaviors of $Z_N^{ib}$
and $Z_N^b$, if confirmed, would also identify scaling 
corrections to the leading behavior in Eq.~(1).
The relative magnitudes of the amplitudes
will play a key role in determining up to what extent
these corrections are important at finite $N$. 

In order to proceed we need a meaningful
definition of the boundary and of the interior 
of a collapsed configuration.
First, from now on
we denote as {\it neighbors} the nearest and second neighbors of a site.
For a SAW walk $w$, we define as boundary the set 
of visited sites which are neighbors of at least one non-visited site 
in communication with the exterior (Fig.~\ref{fig:1}). 
In order to communicate with the exterior
such non-visited site must be connected by at 
least one path of empty neighbor sites to the perimeter 
of a large lattice box enclosing the globule.
The interior is then given by the sites visited by the
walk which do not belong to the boundary.
This boundary definition reminds that of percolation cluster 
hull~\cite{Vanderzande,Duplantier_PRL86:branched,DuplantierSaleur_PRL87:_exact_theta}.
We were able to implement it by sampling long chain 
configurations via the nPERM algorithm  (new pruned
enriched Rosenbluth method) with importance
sampling~\cite{Hsu_MN_Grassberger_JCP03:nPERM}. By this algorithm
we evaluated weights proportional to $Z_N$
and to restricted partition functions defined above,
up to $N_{\rm max}=1920$. The nPERM is an extremely efficient tool 
for sampling long compact SAW configurations
and was even applied with success to study native structures of 
lattice proteins~\cite{Hsu_MN_Grassberger_JCP03:nPERM}.
The weight of the subset of
configurations in which the SAW comes back to the origin
gave an estimate of $Z_N^0$.
We explored the collapsed regime at
$\epsilon / T = 0.7$, $0.77$, and $0.85$.
For the case ${\epsilon/T} =0.85$, on which we concentrated most efforts,
we sampled $2\times 10^9$ linear chain configurations for each $N$
($2\times 10^6$ completely independent). 
This took more than one year CPU time on $2$~GHz machines.

\begin{figure}[!t]
\includegraphics[angle=0,width=7.8cm]{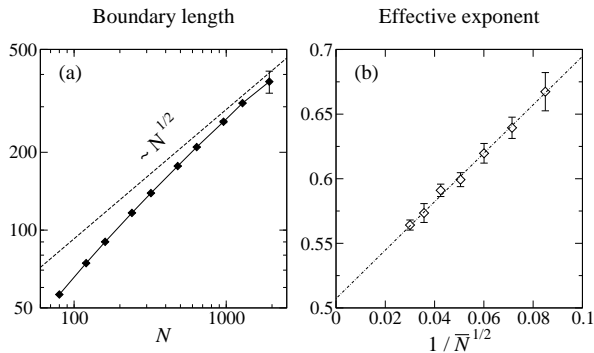}
\caption{(a) $B_N$ vs $N$ for ${\epsilon/T}=0.85$: data
approaches a scaling $\sim N^{1/2}$ (dashed line) for large $N$.
(b) Effective exponent, i.e.\ the slope of
the data in (a), as a function of $\overline N^{-1/2}$.
The line
is a linear fit extrapolating to $\approx 0.5$ for $\overline N\to \infty$.
\label{fig:crust}}
\end{figure}

\begin{figure}[!t]
\includegraphics[angle=0,width=7.2cm]{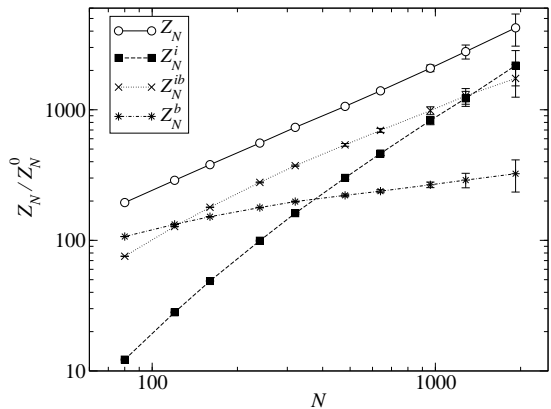}
\caption{
Partition function ratios 
for ${\epsilon/T}=0.85$. 
For $N\lesssim 100$ 
the dominant  contribution to $Z_N$ comes from $Z_N^b$,
for  $100 \lesssim N\lesssim 1300$ $Z_N^{ib}$ dominates, while $Z_N^{i}$ becomes the leading term 
only for $N\gtrsim 1300$. Lines emphasize the trends.
\label{fig:ZZ}}
\end{figure}

\begin{figure}[!bt]
\includegraphics[angle=0,width=7.2cm]{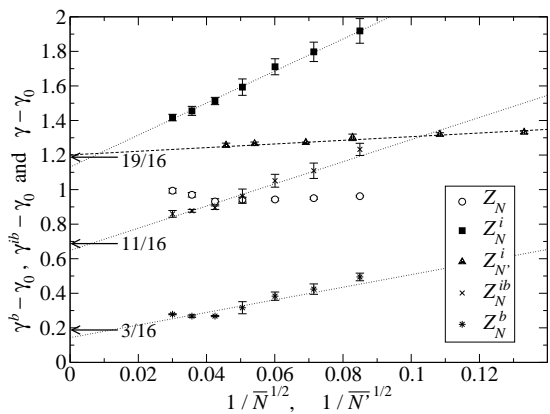}
\caption{Effective $\gamma - \gamma_0$ 
as a function of $\overline N^{-1/2}$ (dotted) or
$\overline{N'}^{-1/2}$ (dashed) for 
$\epsilon/T=0.85$. 
Fits of $\gamma^{ib} - \gamma_0$ 
and of $\gamma^b - \gamma_0$ are also shown.
Arrows mark the DP value $\gamma - \gamma_0=19/16$, and 
$\gamma - \gamma_0 -1/2$ and $\gamma - \gamma_0-1$.
\label{fig:ZZ-expo}}
\end{figure}

We first checked how the average number, $B_N$, of sites on the 
boundary [Fig.~\ref{fig:1}(a)] grows
with $N$. In Fig.~\ref{fig:crust}(a)
we show $B_N$ vs $N$ in a log-log scale for ${\epsilon/T=0.85}$. The data 
give clear evidence that asymptotically $B_N \sim N^{1/2}$. 
Thus, it makes sense to check if the scalings in Eqs.~(\ref{eq:ZZZ})
are consistent with the data, and also provide the leading
scaling correction mechanism to Eq.~(1). We
considered the ratios of all partition functions with
respect to $Z_N^0$. These ratios are reported as a function
of $N$ in the log-log plots of Fig.~\ref{fig:ZZ}, which give an instructive 
representation of the role played by the scaling corrections
at various $N$. The logarithmic slope of
$Z_N/Z_N^0$ vs $N$, which should be $\gamma-\gamma_0$, 
appears to have strong corrections which still sensibly bend the curve 
at the highest values of $N$. Remarkably, the slope of
$Z_N^i/Z_N^0$ approaches nearly the same value in this extremal
region, suggesting that indeed the leading source of corrections is
primarily in $Z_N^{ib}$, as discussed above. The fact that up to
$N \approx 1300$ the curve for $Z_N^i$ remains below 
that for $Z_N^{ib}$ is due to a large
ratio $A_{ib}/A_i$. 
The extrapolated slopes
of the curves are consistent with $\gamma -\gamma_{ib} =
\gamma_{ib} -\gamma_b = 1/2$ according to Eqs.~(\ref{eq:ZZZ})
(see Fig.~\ref{fig:ZZ-expo}).

The above results suggest that the asymptotics of the collapsed 
regime should be best determined by isolating and studying $Z_N^{i}$.  
To confirm this, we evaluated an effective entropic
exponent $\gamma-\gamma_0$ using a weighted linear least square fit of
$\log_{10}(Z_N/Z_N^0)$ vs  $\log_{10}N$ (and similarly for $Z_N^{i}$). 
To quantify the approach to the asymptotic scaling, we consider subsets of
four consecutive points $(N_1,N_2,N_3,N_4)$ and fit their slope 
in Fig.~\ref{fig:ZZ}. The results are plotted as a function of 
$1 / \overline N^{1/2}$  in Fig.~\ref{fig:ZZ-expo}, 
where  $\overline N = (N_1 N_2 N_3 N_4)^{1/4}.$
In the case of $Z_N$ there is much curvature in the plot. 
Quite remarkably in the case of $Z_N^i$ the plot
appears pretty linear and extrapolates very close to
$\gamma -\gamma_0 =
19/16$ for $N \to \infty$.
This linearity clearly indicates that $N^{-1/2}$ represents the 
leading scaling correction. 
One can push further the analogy with DP, imagining that
the $N' = N-B_N$ monomers in the interior of the globule are 
representative of the DP phase, the remaining $B_N$ ones just
forming a kind of box boundary. 
In other words, the partition function  can also be parametrized
by the number $N'$ of monomers in the interior, i.e. $Z_{N'} = Z_N$.  
Repeating the same analysis with $N'$ in place of $N$, we obtain 
effective exponents for $Z_{N'}^i/Z_{N'}^0$ 
that lie on an almost horizontal straight line (Fig.~\ref{fig:ZZ-expo}), 
showing that
the scaling correction amplitude has been drastically reduced.
This allows a more accurate extrapolation of $\gamma -\gamma_0 = 1.20(3)$,  
very close to the value $19/16$ expected for DP.

As a further step we determined individual $\gamma$'s by
directly fitting  $\ln Z_N$ with a function of the form 
\begin{equation}
N \ln \mu + \sqrt{N} \ln \mu_1 + (\gamma-1) \ln N + \ln A + A_1/\sqrt{N}
\label{eq:fit}
\end{equation} 
and $Z_N^0$ with a similar one~\cite{note1}.
The term $A_1/\sqrt{N}$ corresponds to the expected scaling
correction. With or without this term,
we found that upon removal of the data at smaller $N$'s the fits 
are not always stable, again a signal of the presence of a strong 
correction to scaling. On the other hand, with the  $A_1/\sqrt{N}$ term,
the best asymptotic estimate of $\gamma$
is expected when all data are included.
On this basis we estimated $\gamma=1.18(4)$ and $\gamma_0 = 0.02(7)$,
again for $\epsilon/T=0.85$. We also
fitted data 
for ${\epsilon/T}=0.70$. At this temperature one observes 
theta point values of $\gamma$ and $\gamma_0$. This means that
this temperature is still too close to the critical value
$\epsilon/T_\Theta=0.665$~\cite{GrassbergerHegger_JP95:theta2d}
to observe the DP scaling in chains with $N\lesssim 2000$. 
However, for ${\epsilon/T}=0.77$ one finds results 
pretty consistent with those found for $\epsilon/T=0.85$,
confirming the expectation that the collapsed phase
is described by $T$-independent exponents. Our 
determinations are reported in Table I.
\begin{table}[!tb]
\caption{\label{tab:1} Estimated and exact exponents.}
\begin{ruledtabular}
\begin{tabular}{l|ll|ll}
${\epsilon/T}$& 
\multicolumn{2}{c|}{walk: $\gamma$}& 
\multicolumn{2}{c}{polygon: $\gamma_0$}\\
& estim.& exact& estim.& exact\\
\hline
$0.7$ & 1.11(3)  &$8/7\;\footnotemark[1]\simeq 1.14$ & 
-0.15(5) & $-1/7\;\footnotemark[1] \simeq -0.14$ \\
\hline
$0.77$ & 1.20(5)& $19/16\;\footnotemark[2]\simeq 1.19$ & 0.00(5)& $0$
\footnotemark[2]  \\
\hline
$0.85$ & 1.18(4)&$19/16\;\footnotemark[2]\simeq 1.19$ & 0.02(7)& $0$
\footnotemark[2] 
\end{tabular}
\end{ruledtabular}
\footnotetext[1]{Exact entropic exponents of the theta point in 
2D~\protect\cite{DuplantierSaleur_PRL87:_exact_theta}.}
\footnotetext[2]{Exponents of DP in 
2D~\protect\cite{Duplantier_JPA86:dense}, assuming $\gamma_0=0$.}
\end{table}
Previous $\gamma$ estimates without scaling 
corrections~\cite{Nidras_JPA96:lowT} included
$\gamma=1.09(8)$ at $\epsilon/T\simeq0.765$ and
$\gamma=1.11(6)$ at $\epsilon/T\simeq0.788$.
While not testing as deeply the collapsed phase, 
these less sharp determinations
are still compatible with ours.

While the direct fits fully confirm the
difference $\gamma -\gamma_0$ extrapolated above,
consistent with the DP value $19/16$,
$\gamma_0=5/6$, as conjectured in Ref.~\cite{Duplantier_PRL93:comment},
appears definitely excluded. 
The fits suggest that $\gamma_0$
should be a much smaller number, possibly zero. 
The value $\gamma_0 =5/6$ was predicted in Ref.~\cite{Duplantier_PRL93:comment}
by assuming that the globule-solvent interface can be
assimilated to a smooth wall without wedges in the DP model. 
However, the circumstance that the boundary has a fractal 
dimension equal to $1$, as directly verified here, does 
not rule out other possibilities. As we argue below,
one should expect a rough, self-affine boundary. 
The attractive forces guaranteeing the cohesion of the
globule should determine a line tension ($\sim \ln \mu_1$) for the boundary,
which is probably the most important factor controlling
its length. The fluctuations of the
length of the a boundary should have a self-affine geometry,
so that its average width grows
like $B_N^{\zeta} \sim N^{\zeta/2}$, $\zeta <1$ being the 
roughness exponent~\cite{Fisher86}.
In 2D the roughness exponent of a line tension controlled
boundary has $\zeta=1/2$.
One can argue the roughness exponent $\zeta$ of the globule boundary
again on the basis of the leading scaling correction $\sim N^{-1/2}$ identified
in the problem. If the boundary is self-affine, $B_N$
should have a subleading correction $\sim N^{\zeta -1}$, with
positive amplitude, due to the rate of growth with $N$ of the average 
width of the boundary profile.
Thus, in our case we should expect $\zeta=1/2$. This correction is clearly
shown by our plot in Fig.~\ref{fig:crust}(b),
where the effective exponents of the scaling of $B_N$ are plotted as a function
of $\overline N^{-1/2}$: they vary linearly as a function of 
$\overline N^{-1/2}$, with positive slope, and
extrapolate to $0.507(3)$ as $\overline N\to \infty$ 
(dot-dashed line in the figure).
A self-affine curve is not differentiable
and it appears unjustified to represent the globule
boundary as equivalent to a smooth contour
for a DP~\cite{Duplantier_PRL93:comment}.
The theory of DP~\cite{Duplantier_JPA86:dense,DuplantierSaleur_NPB87:dense},
while still valid for the collapsed globule, is not applicable in the form 
appropriate for polymers within smooth boxes. 
Thus the topological argument leading to $\gamma_0 = 5/6$
for a collapsed globule does not hold.  One should expect a smaller
value of $\gamma_0$, as the theory predicts that  $\gamma_0$ is maximal
for a DP with smooth  boundary~\cite{Duplantier_PRL93:comment}.
Indeed, here we  find $\gamma_0\approx 0$, possibly
coinciding exactly with $0$, compatible with the hypothesis that a
collapsed polymer is statistically equivalent to a DP with self-affine
rough boundary.

In summary we gave for the first time solid evidence that
the entropic scaling of a collapsed polymer globule in 2D falls in
the universality class of athermal DP.
The existence of the solvent-globule interface appears crucial in several
respects. Besides giving rise to the strong scaling corrections
which hindered so far the analysis of the problem,
with its nontrivial, self-affine stochastic geometry the interface
also determines an unexpected, close to zero value of the
$\gamma_0$ exponent of collapsed rings.

M.B.\ thanks P.~Grassberger and W.~Nadler for useful discussions, and he
acknowledges support by FWO (Flanders) and
by INFM-PAIS02 in Padova, where this work was started.


\end{document}